\documentclass[10pt,twocolumn,prl,aps,amssymb,amsmath,superscriptaddress]{revtex4}
\include{gyrokinetics-macros}
\usepackage{color,graphicx,ulem,bm}

\newcommand{\mbf}[1]{\mathbf{#1}}

\newcommand{\parl}{\parallel}
\newcommand{\pd}[2]{\frac{\partial #1}{\partial #2}}
\newcommand{\unit}[1]{\mathbf{\hat{#1}}}

\newcommand{\flux}{r}
\newcommand{\phaseavg}[1]{\left<#1\right>_{\Lambda}}
\newcommand{\outerscale}{L}
\newcommand{\lmfp}{\lambda_{\textnormal{mfp}}}

\newcommand{\gyroR}[2][s]{\ensuremath{{\left< #2 \right>}_\mathbf{R_\mathrm{#1}}}}
\renewcommand{\eqref}[1]{Eq.\ (\ref{#1})}

\renewcommand{\gyroR}[1]{\ensuremath{{\left< #1 \right>}}}
\newcommand{\gyroRs}[1]{\ensuremath{{\left< #1 \right>}_{s}}}

\newcommand{\vvol}{d^{3}\mbf{v}}

\newcommand{\vpa}{v_{\parallel}}
\newcommand{\vpe}{v_{\perp}}

\newcommand{\beq}{\begin{equation}}
\newcommand{\eeq}{\end{equation}}
\newcommand{\grad}{\nabla}

\newcommand{\rhoi}{\rho_i}

\begin{document}

\title{Turbulent transport and heating of trace heavy ions in hot, magnetized plasmas}

\author{M.\ Barnes}
\email{mabarnes@mit.edu}
\affiliation{Plasma Science and Fusion Center, Massachusetts Institute of Technology, Cambridge, MA 02138, USA}
\affiliation{Oak Ridge Institute for Science and Education, Oak Ridge, TN 37831, USA}
\author{F.\ I.\ Parra}
\affiliation{Plasma Science and Fusion Center, Massachusetts Institute of Technology, Cambridge, MA 02138, USA}
\author{W. Dorland}
\affiliation{Department of Physics, University of Maryland, College Park, MD 20740, USA}

\begin{abstract}

Scaling laws for the transport and heating of trace heavy ions in low-frequency, magnetized plasma
turbulence are derived and compared with direct numerical simulations.  The predicted dependences
of turbulent fluxes and heating on 
ion charge and mass number are found to agree with numerical results for 
both stationary and differentially rotating plasmas.  Heavy ion momentum transport is found
to increase with mass, and heavy ions are found to be
preferentially heated, implying a mass-dependent ion temperature for very weakly
collisional plasmas and for partially-ionized heavy ions in strongly rotating plasmas.

\end{abstract}



\maketitle

\paragraph{Introduction.}


Heavy ions are present in hot, magnetized plasmas both in laboratory experiments and in nature.
These heavy ions are often trace, i.e., their densities are small enough that they have only a small direct effect on the bulk plasma dynamics.  
Nonetheless, trace heavy ions are important in numerous contexts:  main ion properties are
often inferred from heavy ion measurements because heavy ions radiate more 
readily~\cite{islerPPCF94}; 
accumulation of heavy ions 
leads to dilution and increased radiative energy losses in magnetic confinement 
fusion~\cite{meserveyNF76,tokarNF97}; 
and temperature measurements of minority ions in space and astrophysical plasmas indicate
the existence of a novel heating mechanism~\cite{schmidtGRL80,collierGRL96,kohlSP97}.

Considerable effort has gone into understanding the 
particle transport of trace heavy ions,
or impurities, in the context of magnetized toroidal plasmas for fusion.  
In particular, the scaling with charge number $Z$
and mass number $A$ of the impurity particle
flux were predicted with a quasilinear fluid model and found to be in relatively good agreement
with numerical and experimental results~\cite{angioniPRL06,angioniPPCF09}.
However, little to no work has been done on impurity momentum and energy fluxes or for turbulent
heating of impurities.  The latter may play a role not only in fusion plasmas, but also
in the context of astrophysical plasmas, where the temperature of minority ions has been observed to increase with increasing ion mass~\cite{schmidtGRL80,collierGRL96,kohlSP97}.  Cyclotron heating~\cite{cranmerApJ99} and stochastic heating via large-amplitude fluctuations~\cite{chandranApJ10a} have
been proposed as possible explanations for this
mass dependence.  The turbulent heating mechanism described
here provides an alternative explanation for the mass dependence of the minority ion temperature
that is present even for low frequency, low amplitude fluctuations.

In this Letter we use local, nonlinear, $\delta f$-gyrokinetic 
theory~\cite{cattoPP78,friemanPoF82,howesApJ06} to provide
scaling predictions for trace heavy ion particle, momentum, and energy fluxes, as well as
turbulent heating in hot, magnetized plasmas.  This approach has already proven successful
in determining scalings of temperature-gradient driven turbulence in tokamaks~\cite{barnesPRL11b}.  
We consider an inhomogeneous, axisymmetric plasma rotating toroidally at angular frequency $\omega_{\phi}$, immersed in a curved, inhomogeneous magnetic field.  To simplify our analysis, we
restrict our attention to a region of plasma with rotation speed well below the ion sound speed 
but with a strong rotation gradient.  We also consider only moderate values of $\beta=8\pi p/B^2 \lesssim 1$, where $p$ is the mean plasma pressure and $B$ is the mean magnetic field magnitude.
This is directly applicable to toroidal confinement
experiments in magnetic confinement fusion, but the scaling laws we obtain are general: they
do not change for a stationary, homogeneous plasma slab and therefore also pertain
to various space and astrophysical plasmas.


\paragraph{Gyrokinetic turbulence.}

The $\delta f$-gyrokinetic theory is obtained by performing an asymptotic expansion in the 
small ratio of the Larmor radius, $\rho$, to system size, $\outerscale$, and averaging
over the fast Larmor motion of particles.  It is valid for low-amplitude
turbulence with time scales long compared to the Larmor frequency, $\Omega$,
and spatial scales comparable to $\rho$ and $\outerscale$ in the directions across and
along the mean magnetic field, respectively.  While initially developed for magnetic
confinement fusion plasmas, $\delta f$-gyrokinetics can also be applied to
small-scale turbulence in the solar wind, solar corona, accretion disks, and galaxy 
clusters ~\cite{howesJGR08,schekApJ09}.

We use ($\mbf{R},\mu,\varepsilon$) as our coordinate system, where $\mbf{R}$ is the position of
the center of a particle's Larmor orbit, $\varepsilon=m v^2/2$ its kinetic energy, 
and $\mu=m \vpe^2/2B$ 
its magnetic moment, with $m$ its mass and $v$ its speed.  The subscripts $\perp$ and $\parl$
are used to denote the components perpendicular and parallel to the mean magnetic field,
respectively, with the magnetic field magnitude given by $B$.
With this choice of coordinates, the electromagnetic gyrokinetic equation governing
the evolution of the fluctuating piece of the distribution function, $\delta f_s$, is
\beq
\begin{split}
\frac{Dg_s}{Dt} &+  \dot{\mbf{R}}_s \cdot \grad \left(g_s 
+ \frac{Z_s e\gyroRs{\chi}}{T_s} F_{M,s}\right) - \big<C[\delta f_s]\big>_s\\
&= -\gyroRs{\mathbf{v}_{\chi}} \cdot \left(\grad F_{M,s} 
 + R\grad\omega_{\phi}\frac{m_s\vpa}{T_s}F_{M,s}\right),
\end{split}
\label{eqn:gk}
\eeq
where $g_s = \delta f_s + Z_seF_{M,s}(\Phi-\gyroRs{\chi})/T_s$, 
$\gyroRs{.}$ denotes an average over Larmor angle at fixed
$\mbf{R}_s$, $\gyroRs{\chi}=\gyroRs{\Phi-\vpa\delta A_{\parl}/c
+\int_0^{\mu_s} d\mu_s' \delta B_{\parl}/Z_se}$~\footnote{The fields $\Phi$, $\delta A_{\parl}$, 
and $\delta B_{\parl}$ are independent of Larmor angle at 
fixed particle position, $\mbf{r}$,
but not at fixed $\mbf{R}=\mbf{r}+\mbf{v}_{\perp}\times\unit{b}/\Omega$.  Thus care must be taken
to specify which spatial coordinate is held fixed for velocity integration.  The $\mu$-integral
contained in $\gyroR{\chi}$ is performed at fixed $\mbf{R}$, but all other velocity integrals
in this Letter are performed at fixed $\mbf{r}$.}, 
$\Phi$ is the fluctuating electrostatic potential,
$\delta A_{\parl}$ and $\delta B_{\parl}$ are the parallel components of the 
fluctuating magnetic vector potential and magnetic field, respectively,
$Z_s$ is the charge number, $e$ the proton charge, 
$c$ the speed of light, $T_s$ the mean temperature,
$F_{M,s}$ is a stationary Maxwellian distribution of velocities 
in the frame rotating with velocity $\mathbf{u}=R^2\omega_{\phi} \grad\phi$,
$\phi$ is the toroidal angle, $R$ the plasma major radius,
$D/Dt = \partial/\partial t + \mbf{u}\cdot \grad$,
$\dot{\mbf{R}}_s=\mathbf{v}_{\parallel}+\mathbf{v}_{M,s} 
+ \gyroRs{\mathbf{v}_{\chi}}$, with 
$\mbf{v}_{M,s}=\unit{b}/\Omega_s \times (\vpa^2 \unit{b}\cdot \grad\unit{b}+\vpe^2 \grad B/2B)$ the drift velocity due to a curved, inhomogeneous mean magnetic
field and $\gyroRs{\mbf{v}_{\chi}}=c \unit{b}\times\grad\gyroRs{\chi}/B$ the drift due 
to the fluctuating electromagnetic fields, 
$\unit{b}$ the unit vector along the mean magnetic field,
$\Omega_s=Z_s e B/m_s c$ the Larmor frequency, 
and $C$ describes two-particle Coulomb interactions.
Plasma species is indicated by the subscript $s$, which we
henceforth drop unless it is needed to avoid ambiguity.

By definition, the trace ions considered here do not contribute to the fields.  
They are instead determined solely
by the electron and main ion dynamics through the low-frequency Maxwell's equations,
supplemented by the quasineutrality constraint:
\begin{align}
0 &=\sum_s Z_s \int \vvol \ \delta f_{s},
\label{eqn:QN} \\
\grad_{\perp}^2 \delta A_{\parl} &= -\frac{4\pi}{c}\sum_s Z_s e \int \vvol \ \vpa \delta f_{s},
\label{eqn:apar} \\
\grad_{\perp}\delta B_{\parl} &= \frac{4\pi}{c}\sum_s Z_s e \int \vvol \left(\unit{b}\times\mbf{v}_{\perp}\right) \delta f_{s},
\label{eqn:bpar}
\end{align}
where $\Phi$ enters Eqs.~(\ref{eqn:QN}-\ref{eqn:bpar}) through the definition for $\delta f$
given below Eq.~(\ref{eqn:gk}).

With $g$ and $\{\Phi,\delta A_{\parl},\delta B_{\parl}\}$ specified by 
Eqs.~(\ref{eqn:gk}-\ref{eqn:bpar}), one can evaluate the turbulent heating,
\beq
H \equiv Ze\phaseavg{\chi\left(\left(\mbf{v}_{\parallel}+\mbf{v}_M\right)\cdot \nabla g-\gyroR{C[\delta f]}\right)} - \Pi \pd{\omega_{\phi}}{\flux},
\label{eqn:heating}
\eeq
and the turbulent fluxes,
\begin{align}
\Gamma &= \bigg< \delta f \gyroR{\mbf{v}_{\chi}}\cdot\grad\flux\bigg>_{\Lambda},
\label{eqn:pflx}\\
Q &=  \bigg<\varepsilon \ \delta f \gyroR{\mbf{v}_{\chi}}\cdot\grad\flux\bigg>_{\Lambda},
\label{eqn:qflx}\\
\begin{split}
\Pi &= 
m R^2 \bigg<\delta f(\mbf{v}\cdot\grad\phi)\gyroR{\mbf{v}_{\chi}}\cdot\grad\flux\bigg>_{\Lambda}\\
&- \frac{Ze}{c}R^2 (\unit{b}\cdot\nabla\phi)\bigg<\delta f(\mbf{v}_{\perp}\cdot\grad\flux)\delta A_{\parl}\bigg>_{\Lambda},
\end{split}
\label{eqn:vflx}
\end{align}
where $\flux$ labels surfaces of constant mean pressure, 
$\phaseavg{a}=\int d^3 \mbf{r} \int \vvol \ a/\int d^3 \mbf{r}$ is an integral over all velocity space
and over a volume of width $w$ ($\rho \ll w \ll \outerscale$) encompassing the mean magnetic field
line of interest,
and $\Gamma$, $\Pi$, and $Q$ are the particle, toroidal
angular momentum, and energy fluxes, respectively.  Note that the momentum flux defined in 
Eq.~(\ref{eqn:vflx}) does not include each species' contribution to the Maxwell stress.

\begin{table}[t]
\caption{Scalings, $S$, for turbulent fluxes and heating}
\centering
\begin{tabular}{ c | c | c | c |}
\\ [-2.5ex]
& $\ \ \left|\dfrac{d\omega_{\phi}}{dr}\right| \sim \dfrac{Z}{A} \dfrac{v_{ti}}{R^2}$ \ \ & \ \ $\left|\dfrac{d\omega_{\phi}}{dr}\right|\ll \dfrac{Z}{A} \dfrac{v_{ti}}{R^2} \ \ $ & $\ \ \left|\dfrac{d\omega_{\phi}}{dr}\right|\gg \dfrac{Z}{A}\dfrac{v_{ti}}{R^2} \ \ $ \\ [1.8ex]
\hline
$g_0$ & $A^{1/2}$ or $Z/A^{1/2}$ & $Z/A^{1/2}$ & $A^{1/2}$ \\
$g_1$ & 1 or $Z/A$ & 1 or $Z/A$ & 1 \\
$\Gamma$ & 1 or $Z/A$& $1$ or $Z/A$ & 1 \\
$Q$ & 1 or $Z/A$& $1$ or $Z/A$ & 1 \\ 
$\Pi$ & $A$ or $Z$ & $Z$ & $A$ \\
$H$ & $Z^2/A$, $A$, or $Z$ & $Z^2/A$ & $A$ or $Z$ \\ [0.5ex]
\hline
\end{tabular}
\label{table:scalings}
\end{table}

\paragraph{Expansion in $A^{1/2}$.}  To obtain scaling laws for the
turbulent fluxes and heating of trace heavy ions, we take $Z\sim A\gg 1$, 
$d\omega_{\phi}/dr\sim v_{ti}/R^2$, and
expand $g=g_0+g_1+...$ in powers of $A^{1/2}$.  Here $v_{ti}$ is the main ion thermal speed.  
We restrict our attention to $\beta=8\pi p/B^2\lesssim 1$,
and assume the collisional mean free path is
sufficiently long that collisions may be neglected in our analysis.  
In what follows, we keep $Z$ and $A$ dependences separate
so that we can consider the subsidiary expansion $A^{1/2} \ll Z \ll A$.

Because the heavy ions
are trace, their space and time scales are those of the bulk plasma turbulence.  Thus,
$Z$ and $A$ only enter Eq.~(\ref{eqn:gk}) through explicit factors of $m$, $v\sim v_{t}$, and $Z$,
as well as through $g$ itself.  In what follows, we assume the ratio of the heavy ion to proton
temperature is much smaller than $A$, giving $v_t \sim A^{-1/2}$.  
The two lowest order equations in our expansion are thus
\begin{align}
\begin{split}
&\frac{Dg_0}{Dt} + \gyroR{\mbf{v}_E}\cdot\grad g_0 = -\frac{Z e}{T}F_M \mbf{v}_{\parl} \cdot \grad \gyroR{\Phi}\\
& \ \ \ \ \ \ \ \ \ \ \ \ \ \ \ \ \ \ \ \ \ \ \ -\frac{m \vpa}{T}F_{M}\gyroR{\mbf{v}_E}\cdot R\grad\omega_{\phi},
\label{eqn:g0}
\end{split}
\\
\begin{split}
&\frac{Dg_1}{Dt} + \gyroR{\mbf{v}_E}\cdot\grad g_1
+ \left(\mbf{v}_{\parl}+\gyroR{\mbf{v}_{A}}\right)\cdot \grad g_0 \\
&= -\frac{ZeF_M}{T}\left(\mbf{v}_M\cdot\grad \gyroR{\Phi} 
+\frac{\vpa^2}{c}\unit{b}\cdot\grad\gyroR{\delta A_{\parl}}\right)\\
&\ \ \ \ -\frac{m\vpa}{T}F_M\gyroR{\mbf{v}_A}\cdot R\grad\omega_{\phi}
-\gyroR{\mbf{v}_E}\cdot\grad F_M,
\label{eqn:g1}
\end{split}
\end{align}
where $\mbf{v}_E=c\unit{b}\times\grad\Phi/B$ and 
$\mbf{v}_{A}=\vpa\unit{b}\times\grad\delta A_{\parl}/B$.    

There are two possible scalings for both $g_0$ and $g_1$ due to a competition between
terms with different $A$ and $Z$ dependences in Eqs.~(\ref{eqn:g0}) and~(\ref{eqn:g1}).  
In particular, $g_0 \propto A^{1/2}, \ Z/A^{1/2}$, and $g_1 \propto 1, \ Z/A$.
By considering the limits $|d\omega_{\phi}/dr|\ll (Z/A)v_{ti}/R^2$ and 
$|d\omega_{\phi}/dr|\gg (Z/A)v_{ti}/R^2$, the number of such scalings is reduced.
The $A$- and $Z$-scalings for $g_0$ and $g_1$ in these limits, as well as for the general
case, are summarized in Table~\ref{table:scalings}.


\paragraph{Flux and heating scalings.}

If $g_0(\vpa)$ is a solution to Eq.~(\ref{eqn:g0}), then $-g_0(-\vpa)$ is also a solution.  Thus, $\overline{\int_{-\infty}^{\infty} d\vpa g_0\{\Phi,\delta A_{\parl},\delta B_{\parl}\}}=0$, where the overline denotes a statistical average.
As a result, $g_0$ does not contribute to the lowest order (i.e., electrostatic) heating or particle 
and heat fluxes, Eqs.~(\ref{eqn:heating})-(\ref{eqn:qflx}), whose integrands are otherwise 
even  functions of $\vpa$.  Conversely, the lowest order momentum flux integrand has a 
component proportional to $m \vpa$, so $\Pi \sim m v_{t} g_0 \propto A^{1/2} g_0$.  
Using our scalings for $g_0$, we see that $\Pi$
has competing terms scaling as $Z$ and $A$, respectively.

Note that Eq.~(\ref{eqn:g1}) has a $\vpa$ symmetry that is opposite that of Eq.~(\ref{eqn:g0}): 
if $g_1(\vpa)$ is a solution, then  $g_1(-\vpa)$ is also a solution.  
Furthermore, for all higher order equations, one can show that
the symmetry in $\vpa$ alternates between that of Eqs.~(\ref{eqn:g0}) and~(\ref{eqn:g1}).
As a result, the only components of $g$ that contribute to the particle and heat fluxes and heating are
$g_1$, $g_3$, etc.  Using Eqs.~(\ref{eqn:pflx}) and~(\ref{eqn:qflx}), we have $\{\Gamma,Q\}\sim g_1$,
which in the general case has competing terms scaling as $Z/A$ and $1$ (no $Z$ or $A$
dependence), respectively.  

The first term in the heating expression, (\ref{eqn:heating}), is the Joule heating and is scaled up by an
explicit factor of $Z$ (arising from the current), while the second term is viscous heating.  
At lowest order, the Joule heating term gives $H \propto (Zv_{\parl}g_0, \ Z g_1)$, giving
$H \propto (Z, \ Z^2/A)$.  The viscous heating is proportional to $\Pi\propto (Z, \ A)$.
$H$ thus has competing terms scaling as $Z^2/A$, $Z$, and $A$, respectively.
The scalings of the various fluxes
and heating are summarized in Table~\ref{table:scalings}.

\begin{figure}
\includegraphics[height=2.2in]{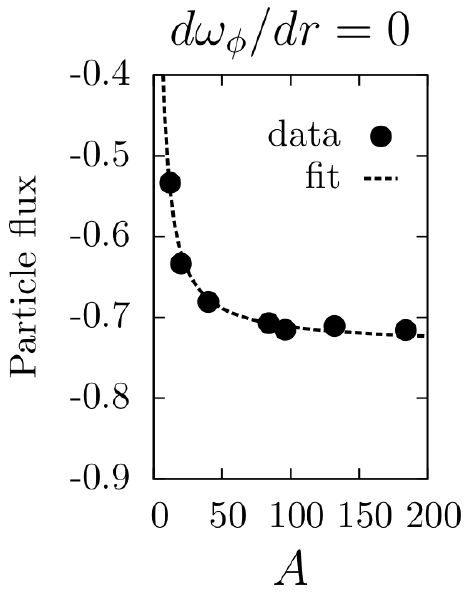}
\includegraphics[height=2.2in]{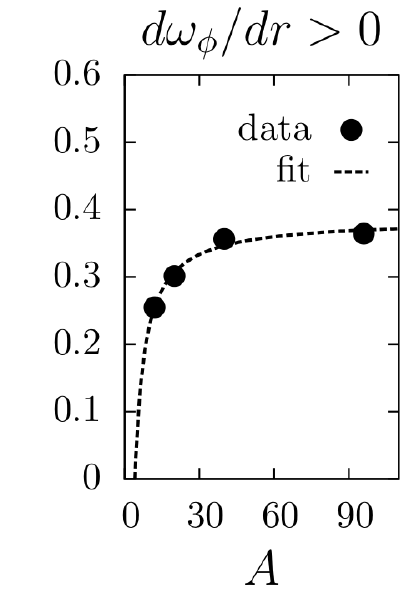}
\caption{Normalized particle flux, $(\Gamma_s/n_s v_{ti})(\outerscale/\rhoi)^2$, vs.~mass number,
$A$, for cases with and without differential rotation, $\omega_{\phi}$.  The dashed line is
a least-squares fit using our scaling predictions, given by $-0.7+2.3/A$ and $0.4-1.6/x$ 
for the left and right plots, respectively.}
\label{fig:pflx}
\end{figure}

\paragraph{Minority ion temperature.}

Integrating Eq.~(\ref{eqn:heating}) by parts in time and using Eq.~(\ref{eqn:gk}),
the heating can be expressed as~\cite{krommesPoP94,howesApJ06,abelRPP12}
\beq
\begin{split}
H_s &= -\phaseavg{ \frac{T_s \delta f_s}{F_{M,s}} \ C[\delta f_s]} \\
&+ \left(Q_s - \frac{3}{2}\Gamma_s\right) \pd{\ln T_s}{r} + \Gamma_s \pd{\ln n_s}{r}.
\end{split}
\label{eqn:heating2}
\eeq
Our scalings indicate that $H_s$ increases in magnitude with $A$ or $Z$, but $\Gamma_s$
and $Q_s$ do not.  The first term in Eq.~(\ref{eqn:heating2}) must thus dominate for $A$ or $Z$
large.  
This term, which we identify as the collisional entropy generation, is positive definite
when summed over species.  We argue that it is also positive species by species
for the low collisionalities considered here.

\begin{figure}
\includegraphics[height=2.2in]{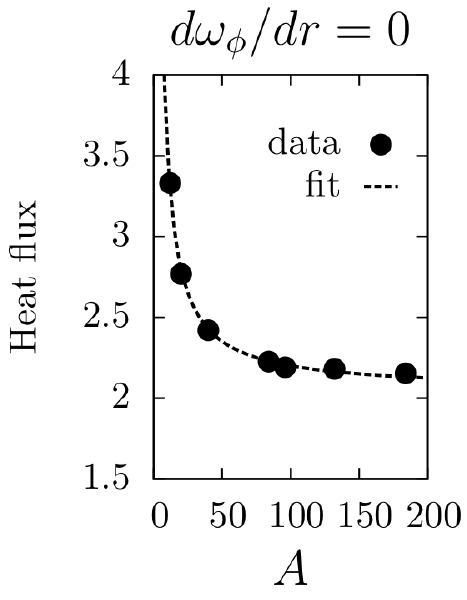}
\includegraphics[height=2.2in]{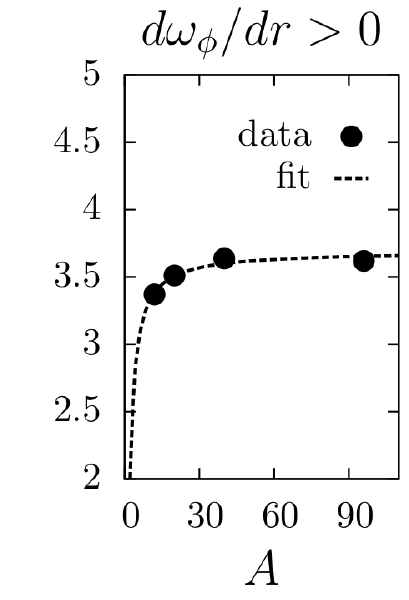}
\caption{Normalized heat flux, $(Q_s/n_s T_i v_{ti})(\outerscale/\rhoi)^2$, vs.~mass number,
$A$, for cases with and without differential rotation, $\omega_{\phi}$.  The dashed line is
a least-squares fit using our scaling predictions, given by $2.0+15/A$ and $3.7-3.7/A$ 
for the left and right plots, respectively.}
\label{fig:qflx}
\end{figure}

The collision operator, $C$, consists of a test-particle piece, which is a diffusion 
operator in velocity space, and a field-particle piece, which is an integral operator~\cite{helander}.  Both contributions are inversely proportional to the collisional mean free path and thus small,
except at small scales in the velocity space where large derivatives in the test-particle operator
compensate~\cite{abelPoP08,barnesPoP09,schekApJ09}.  The test-particle operator should 
thus dominate in weakly collisional plasmas, and its diffusive nature ensures that its contribution 
to entropy generation is positive-definite.

Consequently, trace heavy ions must be heated by turbulence instead of cooled.
For this heating process to subside, the trace ion temperature must become large enough
to interfere with our large $A$ expansion.  This happens when the heavy ion temperature
exceeds the main ion temperature by a factor of $A \sim Z$.  In this limit, the turbulent heating
$H$ becomes comparable to the heat flux $Q$ so that $H$ is no longer required to be positive
definite.
Our theory thus predicts that heavy ions will be hotter than light ions by a factor of $A\sim Z$
-- but only if turbulent heating is larger than collisional temperature equilibration.  

The collisional temperature equilibration of the main ions, $i$, 
and a trace heavy ion species, $s$, is $\mathcal{E}_s\equiv (8/3\sqrt{\pi})(Z_s^2/A_s)n_s \Delta T_s v_{ti}/\lambda_{\textnormal{mfp}}$,
where $\Delta T_s = T_s-T_i$, and $\lmfp$ is the mean free path for collisions between the main ions.
From Eq.~(\ref{eqn:heating}), we estimate $H_s\sim S n_s T_i (\delta n_i/n_i)^2 v_{ti}/L$,
where $S$ is the scaling of $H$ with $A$ and $Z$ given in Table~\ref{table:scalings}, and
we have assumed $e\Phi/T_i \sim \delta n_i / n_i \sim \delta n_s / n_s$.
The ratio of turbulent heating to collisional temperature equilibration is thus
$H/\mathcal{E}\sim S (A/Z^2) (\lmfp/\outerscale) (\delta n_i/n_i)^2\sim S(A/Z^2) \lmfp/v_{ti}\tau_E$,
with $\tau_E$ the characteristic time scale over which the equilibrium density and temperature
vary.

\begin{figure}
\includegraphics[height=2.2in]{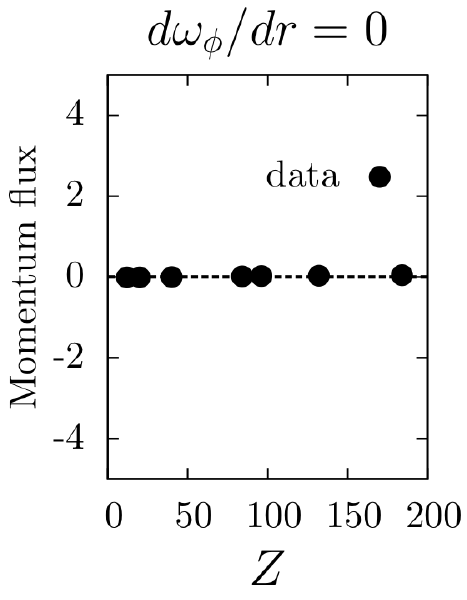}
\includegraphics[height=2.2in]{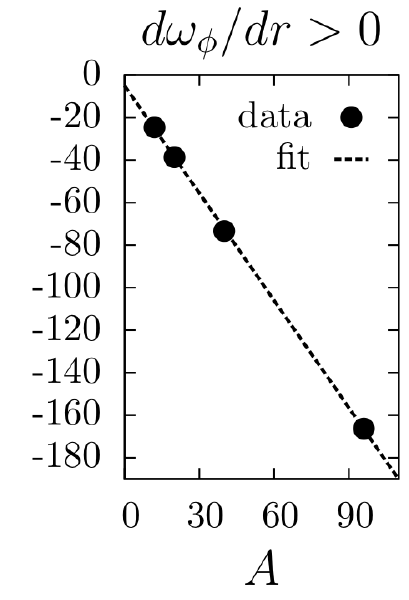}
\caption{Normalized toroidal angular momentum flux, 
$(\Pi_s/m_i n_s \outerscale v_{ti}^2)(\outerscale/\rhoi)^2$, vs.~charge number, $Z$, and mass number,
$A$, for cases with and without differential rotation, $\omega_{\phi}$.  The dashed lines are
least-squares fits using our scaling predictions, given by $0$ and $-1.7A-5.1$ 
for the left and right plots, respectively.  The fact that $\Pi=0$ for the case with 
$d\omega_{\phi}/dr=0$ is a consequence of a symmetry property of the gyrokinetic 
equation~\cite{parraPoP11}.}
\label{fig:vflx}
\end{figure}

\paragraph{Numerical results.}

To test our predictions for the scalings of turbulent transport and heating, we employ
the local, $\delta f$-gyrokinetic code \texttt{GS2}~\cite{dorlandPRL00}.  We consider
an axisymmetric system with sheared magnetic field lines mapping out nested toroidal surfaces 
with circular cross sections (known as the Cyclone Base Case~\cite{dimitsPoP00}
and parametrized using the Miller local equilibrium model~\cite{millerPoP98}).  Each simulation is electrostatic and includes kinetic electrons, as well as kinetic
main and trace heavy ions with a wide range of $Z$ and $A$ values.  The turbulence is
driven by gradients in the mean ion and electron densities and temperatures, with 
$R_0 (d\ln n/dr) = 2.2$ for the electrons and main ions, and $R_0 (d\ln T/dr)=6.9$ for all species,
with $R_0$ the major radius at the center of the constant pressure surface.  The collision
frequency is chosen small, $R_0/\lmfp=0.003$, so that heavy ion collisions do not
affect our scalings.

\begin{figure}
\includegraphics[height=2.2in]{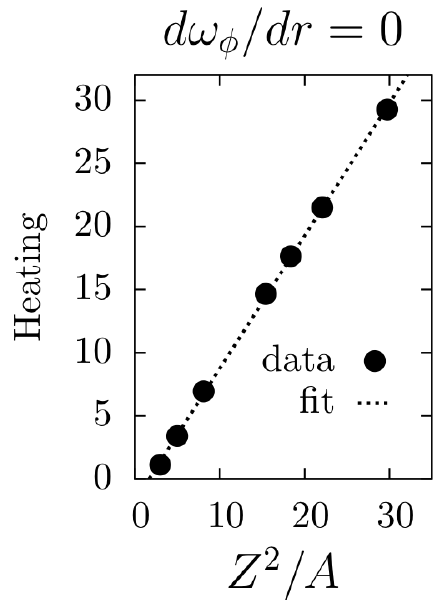}
\includegraphics[height=2.2in]{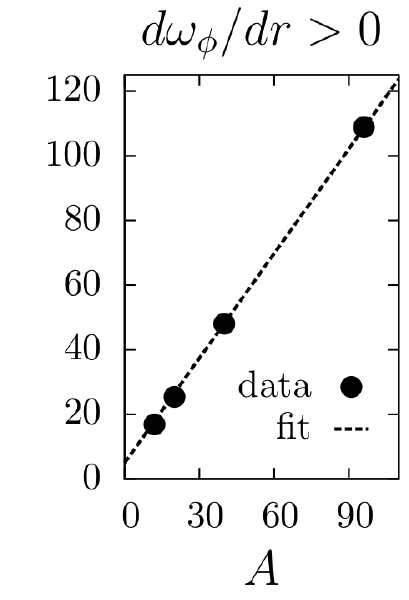}
\caption{Normalized heating, $H_s(\outerscale/n_s T_i v_{ti})(\outerscale/\rhoi)^2$, 
vs.~charge number, $Z$, and mass number,
$A$, for cases with and without differential rotation, $\omega_{\phi}$.  The dashed lines are
least-squares fits using our scaling predictions, given by $1.1Z^2/A-1.8$ and $1.1A+5.0$ 
for the left and right plots, respectively.}
\label{fig:heating}
\end{figure}

Two sets of simulations were carried out: one with a stationary plasma ($d\omega_{\phi}/dr=0$)
and one with a differentially rotating plasma ($d\omega_{\phi}/dr=4.67 v_{ti}/R^2$).  
The simulation results are shown in Figs.~(\ref{fig:pflx})-(\ref{fig:heating}).  Data points for fluxes
and heating at various $Z$ and $A$ values are plotted as solid circles and fit
using a least-squares analysis with the predicted lowest order $Z$ and $A$ dependences, 
as well as the first order correction.  In each case, the predicted scalings fit the data well.
It should be noted that the momentum flux for $d\omega_{\phi}/dr=0$ is zero for all species
due to a fundamental symmetry of the $\delta f$-gyrokinetic equation~\cite{parraPoP11}.

\paragraph{Discussion.}

We now discuss the implications of the trace heavy ion scalings derived in this Letter.
First, the preferential heating of heavy ions should lead to large temperature disparities
between different ion species in nearly collisionless plasmas.  Many space and astrophysical
plasmas are weakly collisional enough (i.e., $(\delta n/n)^2 > \outerscale/\lmfp$)
that turbulent heating should dominate over collisional equilibration, and preferential
heating of heavy ions is indeed observed~\cite{schmidtGRL80,collierGRL96}.  
However, for such low collisionalities
the equilibrium can deviate strongly from the isotropic Maxwellian assumed in our analysis,
which cannot consequently address the large $T_{\perp}/T_{\parallel}$ values
observed in coronal holes and the fast solar wind~\cite{kohlSP97}.

Magnetic confinement fusion 
plasmas typically have $(\delta n/n)^2 < \outerscale/\lmfp$ so that collisional temperature equilibration dominates over turbulent heating and all ions have the same temperature.  
However, for rotating plasmas our results indicate that the turbulent heating is enhanced by
an additional factor of $(A/Z)^2$ relative to the equilibration.
It may therefore be possible for heavy, partially ionized impurities to be heated by turbulence to temperatures significantly larger than the main ions.

Because the momentum transport of heavy ions is enhanced by $A$, it can
generate flows of order the ion thermal speed for densities as small as $n_i/A$. 
In this limit, heavy ions could thus significantly alter bulk plasma momentum transport.

We thank S. C. Cowley, E. Quataert, and A. A. Schekochihin for useful discussions.
M.B. was supported by a US DoE FES Postdoctoral Fellowship, F.I.P. was supported by US DoE Grant
No DE-FG02-91ER-54109, and computing time was provided by HPC-FF (J\"ulich).


\end{document}